%% file: main.tex
\documentclass[sigconf]{acmart}
\copyrightyear{2026}
\acmYear{2026}
\setcopyright{cc}
\setcctype{by}
\acmConference[FORGE '26]{2026 IEEE/ACM Third International Conference on AI Foundation Models and Software Engineering}{April 12--13, 2026}{Rio de Janeiro, Brazil}
\acmBooktitle{2026 IEEE/ACM Third International Conference on AI Foundation Models and Software Engineering (FORGE '26), April 12--13, 2026, Rio de Janeiro, Brazil}
\acmPrice{}
\acmDOI{10.1145/3793655.3793720}
\acmISBN{979-8-4007-2477-0/2026/04}

\usepackage{multirow}
\usepackage{listings}
\begin{document}

%%
%% The "title" command has an optional parameter,
%% allowing the author to define a "short title" to be used in page headers.
% \title{The Name of the Title Is Hope}
\title{\textsc{DynamicsLLM}: a Dynamic Analysis-based Tool for Generating Intelligent Execution Traces Using LLMs to Detect Android Behavioural Code Smells}

%%
%% The "author" command and its associated commands are used to define
%% the authors and their affiliations.
%% Of note is the shared affiliation of the first two authors, and the
%% "authornote" and "authornotemark" commands
%% used to denote shared contribution to the research.
\author{Houcine Abdelkader CHERIEF}
\orcid{0009-0007-4347-1872}
\affiliation{%
  % \institution{École de technologie supérieure (ÉTS)}
  \institution{École de technologie supérieure}
  \city{Montréal}
  \state{Québec}
  \country{Canada}
}

\author{Florent AVELLANEDA}
\orcid{0000-0003-1030-5388}
\affiliation{%
  % \institution{Université du Québec à Montréal (UQÀM)}
  \institution{Université du Québec à Montréal}
  \city{Montréal}
  \state{Québec}
  \country{Canada}
}

\author{Naouel MOHA}
\orcid{0000-0001-9252-9937}
\affiliation{%
  % \institution{École de technologie supérieure (ÉTS)}
  \institution{École de technologie supérieure}
  \city{Montréal}
  \state{Québec}
  \country{Canada}
}

%%
%% By default, the full list of authors will be used in the page
%% headers. Often, this list is too long, and will overlap
%% other information printed in the page headers. This command allows
%% the author to define a more concise list
%% of authors' names for this purpose.
% \renewcommand{\shortauthors}{Trovato et al.}

%\makeatletter
%% Commands for tools
%\newcommand{\tool}[1]{\textsc{#1}}
%\newcommand{\ourtool}{\textsc{DynamicsLLM}}
% \newcommand{\appsnumber}{333}
%\newcommand{\validatednumber}{136}
%\newcommand{\ygg@basicalert}[2]{{\bfseries\sffamily\scriptsize#1}{\sf\small$\blacktriangleright$\textit{#2}$\blacktriangleleft$}}
%\newcommand{\Naouel}[1]{\ygg@basicalert{Naouel}{\textcolor{red}{\r#1}}}
%\newcommand{\FA}[1]{\ygg@basicalert{Florent}{\textcolor{blue}{#1}}}
%\newcommand{\FAA}[2]{\ygg@basicalert{Florent}{\textcolor{blue}{\sout{#1} #2}}}
%\newcommand{\Abdelkader}[1]{\ygg@basicalert{Abdelkader}{\textcolor{green}{\r#1}}}
%\newcommand{\todo}[1]{\ygg@basicalert{To update}{\textcolor{red}{\r#1}}}
%\makeatother

%%
%% The abstract is a short summary of the work to be presented in the
%% article.
\input{Sections/0_abstract}

%\ccsdesc[500]{Software and its engineering~Software reliability}

%%
%% Keywords. The author(s) should pick words that accurately describe
%% the work being presented. Separate the keywords with commas.
%% A "teaser" image appears between the author and affiliation
%% information and the body of the document, and typically spans the
%% page.
% \begin{teaserfigure}
%   \includegraphics[width=\textwidth]{sampleteaser}
%   \caption{Seattle Mariners at Spring Training, 2010.}
%   \Description{Enjoying the baseball game from the third-base
%   seats. Ichiro Suzuki preparing to bat.}
%   \label{fig:teaser}
% \end{teaserfigure}

% \received{20 February 2007}
% \received[revised]{12 March 2009}
% \received[accepted]{5 June 2009}

%%
%% This command processes the author and affiliation and title
%% information and builds the first part of the formatted document.
\maketitle

% Define commands to highlight text with specific background colors
% \newcommand{\Completed}[1]{\begin{tcolorbox}[colback=green!30,boxrule=0pt,arc=0pt,auto outer arc,left=1pt,right=1pt,top=1pt,bottom=1pt]\strut #1\end{tcolorbox}}
% \newcommand{\Inprogress}[1]{\begin{tcolorbox}[colback=yellow!30,boxrule=0pt,arc=0pt,auto outer arc,left=1pt,right=1pt,top=1pt,bottom=1pt]\strut #1\end{tcolorbox}}
% \newcommand{\Noprogress}[1]{\begin{tcolorbox}[colback=red!30,boxrule=0pt,arc=0pt,auto outer arc,left=1pt,right=1pt,top=1pt,bottom=1pt]\strut #1\end{tcolorbox}}

\section{Introduction}
\input{Sections/1_introduction}

\section{Background}
\input{Sections/2_background}

\section{Related Work}
\input{Sections/3_related_works}

\section{The \textsc{DynamicsLLM} Tool}
\input{Sections/4_tool}

\section{Evaluation}
\input{Sections/5_evaluation}

\section{Conclusion}
\input{Sections/6_conclusion}

\bibliographystyle{ACM-Reference-Format}
\bibliography{main}

%%
%% If your work has an appendix, this is the place to put it.
\appendix

\end{document}

%% file: Sections/0_abstract.tex
\begin{abstract}
    Mobile apps have become essential of our daily lives, making code quality a critical concern for developers. 
    Behavioural code smells are characteristics in the source code that induce inappropriate code behaviour during execution, which negatively impact software quality in terms of performance, energy consumption, and memory.

    \noindent \textsc{Dynamics}, the latest state-of-the-art tool-based method, is highly effective at detecting Android behavioural code smells. %However, it depends on dynamic analysis, necessitating high coverage of code smell-related events to examine execution traces and identify code smells.
    While it outperforms static analysis tools, it suffers from a high false negative rate, with multiple code smell instances remaining undetected.
    Large Language Models (LLMs) have achieved notable advances across numerous research domains and offer significant potential for generating \emph{intelligent} execution traces, particularly for detecting behavioural code smells in Android mobile applications. By \emph{intelligent execution trace}, we mean a sequence of events generated by specific actions in a way that triggers the identification of a given behaviour.

    \noindent We propose the following three main contributions in this paper:
    (1) \textsc{DynamicsLLM}, an enhanced implementation of the \textsc{Dynamics} method that leverages LLMs to intelligently generate execution traces.
    (2) A novel hybrid approach designed to improve the coverage of code smell-related events in applications with a small number of activities.
    (3) A comprehensive validation of \textsc{DynamicsLLM} on 333{} mobile applications from F-DROID, including a comparison with the \textsc{Dynamics} tool. Our results show that, under a limited number of actions, \textsc{DynamicsLLM} configured with \textsc{100\% LLM} covers three times more code smell-related events than \textsc{Dynamics}. The hybrid approach improves LLM coverage by 25.9\% for apps containing few activities. Moreover, 12.7\% of the code smell-related events that cannot be triggered by \textsc{Dynamics} are successfully triggered by our tool.
\end{abstract}

%% file: Sections/1_introduction.tex
\par Mobile applications have become an essential part of our daily lives, with billions of downloads from platforms such as the Apple App Store \cite{appstores2022} and Google Play Store \cite{googleplay2022} every year.
 Ensuring code quality is essential to guarantee efficient, secure and maintainable applications (apps). Code smells are poor software design and implementation practices resulting from bad implementation or design choices \cite{Fowler1999}. They are not necessarily bugs, but they can slow down development, increase the risk of errors and make the code harder to maintain.
Therefore, detecting code smells within software systems is an important priority to decrease technical debt \cite{Dynamics}. Besides common object-oriented code smells \cite{behavioural_code_smells_impirical_study}, mobile apps have their own smells due to resource limitations and constraints, such as memory, performance, and energy consumption. Some of these are behavioural code smells, which are characteristics in the source code that cause inappropriate behaviour during execution. The term “behaviour” specifically refers to execution behaviour, meaning an occurrence or sequence of observable code events or actions during execution \cite{behavioural_code_smells_impirical_study}.

\par To address this problem, many tools exist to detect behavioural code smells in mobile apps, based on static analysis techniques like \textsc{ADoctor} \cite{ADOCTOR} and \textsc{Paprika} \cite{Paprika}. These tools analyse the app under test (AUT) without execution to detect code smells, and many of the reported instances are not really code smells \cite{behavioural_code_smells_impirical_study}.

\par A dynamic, tool-based approach called \textsc{Dynamics} also exists. It instruments the tested applications to collect runtime execution traces, which are then used to detect behavioural code smells \cite{Dynamics}. This method is more effective than static analysis due to the complexity and dynamic nature of mobile applications, achieving an average precision of 92.8\%. However, the average recall is only 53.4\%.
The effectiveness of \textsc{Dynamics} method depends on the quality of the generated sequence of events. Most dynamic analysis methods use random event generation strategies, resulting in a high number of false negatives \cite{PredRacer} and a low testing coverage \cite{GPTDroid}, and are unable to locate native code quickly \cite{JNFuzzDroid}.

\par Automated GUI (Graphical User Interface) testing tools, such as \textsc{MonkeyRunner} 
\cite{monkey_tool} 
and \textsc{Droidbot} 
\cite{Droidbot}, are widely employed to avoid time-consuming and labour-intensive manual testing. These tools explore apps and perform various actions, such as scrolling and clicking, to obtain an execution trace.
%\Naouel{Cette dernière phrase n'est pas claire.}
% based on program analysis for the application verification to obtain the execution trace.
%However, generating execution traces using these tools results in limited coverage. For example, it is challenging for \textsc{MonkeyRunner} to reach certain activities and trigger specific events solely through random events. Moreover, it is even more difficult, if not impossible, to pass certain activities like login and registration, which require filling out forms with exact data formats. This results in a lack of code coverage, missing relevant events, while producing numerous false negatives, which are code smells that were not detected but should have been.
However, a recently published study \cite{akinotcho2024mobile} indicates that even the most sophisticated techniques achieve only around 30\% coverage for real-world applications. These tools face several challenges, including difficulties in generating \emph{semantically meaningful inputs} (such as correctly formatting passwords in registration forms), and performing GUI actions in the \textit{correct sequence} (e.g., executing actions in the required order to bypass the current activity).

%\par \Naouel{La phrase qui suit est trop longue et j'ai dû mal à la comprendre} 
\par Pre-trained large language models (LLMs) have recently emerged as a breakthrough technology in natural language processing and artificial intelligence, demonstrating exceptional performance across various tasks\cite{LLM_software_testing}.
Numerous research efforts \cite{GPTDroid, Droidbot_GPT, DroidAgent} focus on improving software testing tasks, particularly in generating execution traces, among which LLMs are the most promising ones. The approach of testing mobile apps using LLMs is presented as a Q\&A task, where the agent prompts the LLM for the next action, and the LLM responds with the necessary information to complete the task. This approach aims to create \emph{intelligent execution traces}, which are sequences of events generated by specific actions to trigger the identification of a given behaviour. These traces could help dynamic analysis-based tools better detect behavioural code smells by reducing false negatives.
% by doing high coverage
%An intelligent execution trace refers to a series of actions performed on a mobile app, designed to maximize the identification of behavioral code smells. \Naouel{J'ai eu du mal avec ce paragraphe}
%\Naouel{Quel est concrètement le problème ??}\FA{Proposition: While LLM-based trace generators aim to cover as many activities as possible within mobile applications, these tools have yet to be applied specifically to detect code smells.}
%While LLM-based trace generators aim to cover as many activities as possible within mobile applications, these tools have yet to be applied specifically to detect code smells.
However, the use of these LLM-based trace generators has yet to be applied. 

Therefore, in this study, we address this research gap by improving the effectiveness of the dynamic analysis method by producing intelligent traces for the detection of Android behavioural code smells using LLM-based trace generators. We present the following contributions:

\begin{itemize}
    \item First, we propose a fully automatic tool, called \textsc{DynamicsLLM}, that uses freely accessible and open-source LLMs to generate intelligent execution traces for more efficient and accurate detection of behavioural code smells.
    \item Second, a new hybrid approach was proposed to enhance the performance of LLMs in covering more code smell-related events within a 1-hour timeout especially for mobile application that contains fewer number of activities.
    \item Third, the \textsc{DynamicsLLM} tool is evaluated using multiple metrics (precision, recall, and code smell–related event coverage) on a subset of 333{} mobile applications from the same validation dataset used by \textsc{Dynamics} in the open-source F-DROID repository. We then compare the performance of \textsc{DynamicsLLM} with \textsc{Dynamics}, the latest state-of-the-art tool-based method.  %\textsc{Dynamics} is based on  \textsc{MonkeyRunner}, the most random GUI generation tool used in the dynamic analysis of Android apps. 
    Our results show that, within a limited number of GUI actions, \textsc{DynamicsLLM}, using \textsc{100\% LLM}, covers three times more code smell–related events than \textsc{Dynamics}. The hybrid approach further improves LLM coverage by 25.9\% for applications containing four or fewer activities, which represent 63\% of the dataset. Moreover, 12.7\% of the code smell–related events that cannot be triggered by \textsc{Dynamics} are successfully triggered by \textsc{DynamicsLLM}. The replication is available \cite{replicationPackage} for further research and analysis. It contains more than 1,230 hours of execution traces.
    %Moreover, 18\% of code smell-related events that cannot be triggered with the \textsc{Dynamics} tool are indeed triggered by our tool.
    
\end{itemize}
\par This paper is organized as follows. Section~\ref{sec:Backgroud} presents background information about behavioural code smells and the \textsc{DroidAgent} tool. Section~\ref{sec:RelatedWorks} reviews related work. Section~\ref{sec:SolutionDesign} details our proposed approach for detecting behavioural code smells. Section~\ref{sec:Evaluation} presents our evaluation methodology and results. Finally, Section~\ref{sec:Conclusion} concludes the paper.

%% file: Sections/2_background.tex
\label{sec:Backgroud}
In this section, we define behavioral code smells. Then, we present the \textsc{DroidAgent} tool used in the implementation of our tool.
\subsection{Behavioural Code Smells}
To the best of our knowledge, only seven behavioural code smells have been identified in the literature by \citet{behavioural_code_smells_impirical_study}. We present the definitions provided in \cite{Dynamics,Paprika}, along with the corresponding inappropriate behaviours.

\textbf{Durable WakeLock (DW):} A WakeLock is a mechanism
that allows an app to keep the device awake to complete a task.

\noindent \textbf{Inappropriate Behaviour:} A call to the \textit{acquire} method is not followed by a call to the \textit{release} method. This causes an energy leak.

\textbf{Init OnDraw (IOD):} \textit{onDraw} routines are responsible for updating the GUI of Android apps. These routines are invoked
each time the GUI is refreshed (up to 60 times per second).

\noindent \textbf{Inappropriate Behaviour:} Any extra computational work or excessive memory usage performed in \textit{onDraw} is magnified due to the high frequency of invocation.

\textbf{Heavy Processes (HP):} This can be divided into three code smells: (1) \textit{Heavy AsyncTask (HAS)}, where \textit{AsyncTask} is used to perform short background operations; (2) \textit{Heavy Service Start (HSS)}, where Android Services are used to perform heavy operations; and (3) \textit{Heavy Broadcast Receiver (HBR)}, where a broadcast receiver is used to manage communication with the system or other apps.

\noindent \textbf{Inappropriate Behaviour:} Using \textit{AsyncTask}, Services, or Broadcast Receivers requires invoking multiple methods for starting and callback. Some of these methods are executed on the main UI thread and should not be time-consuming or blocking, as this may lead the system to kill the app.

\textbf{No Low Memory Resolver (NLMR):} The \textit{onLowMemory} method is responsible for reducing the memory usage of a running activity.

\noindent \textbf{Inappropriate Behaviour:} If the method is not implemented or fails to release memory when executed, the Android system may automatically kill the activity’s process to free memory, potentially causing an unexpected termination of the program.

\textbf{HashMap Usage (HMU):} The Android framework provides \textit{ArrayMap} and \textit{SimpleArrayMap} as memory-efficient replacements for \textit{HashMap}. These alternatives trigger less garbage collection without a significant difference in performance for maps containing up to hundreds of entries.

\noindent \textbf{Inappropriate Behaviour:} A \textit{HashMap} is used for a small set of objects, or \textit{ArrayMap} and \textit{SimpleArrayMap} are used for large sets.
\subsection{\textsc{DroidAgent} Tool}
\textsc{DroidAgent} is a LLM-based trace generator, designed with an agent-based architecture consisting of four main LLM-based agents, each performing specific tasks: \textbf{Planner}, \textbf{Actor}, \textbf{Observer}, and \textbf{Reflector}. The \textbf{Actor} and \textbf{Observer} form an "inner" loop, working to accomplish tasks that have been planned by the \textbf{Planner} and later reflected upon by the \textbf{Reflector}, as shown in Figure \ref{fig:DroidAgentMechanisme}. We describe the process as follows:
\begin{figure}[htbp]
    \centering
    \scalebox{0.3}{  
    \centerline{\includegraphics{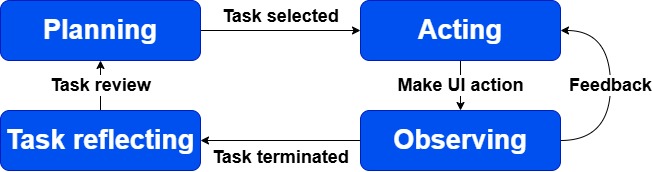}}
    }
    \caption{\textsc{DroidAgent} process}
    \Description{\textsc{DroidAgent} process. The workflow starts with the Planning step, where a task is selected. In the Acting step, the agent performs UI actions. The Observer provides feedback to the Actor, which performs additional UI actions until the task is terminated. The Reflector then provides a task review to the Planner. This process repeats iteratively.}
    \label{fig:DroidAgentMechanisme}
\end{figure}
    \begin{enumerate}
        % Add your items here
        \item \textbf{Planning:} \textsc{DroidAgent} continuously plans high-level tasks to be achieved. These tasks correspond to semantically meaningful steps when testing the app and align with the coherent functionalities of the target application. The \textit{Planner} agent generates viable and diverse tasks while avoiding the repetition of impossible, irrelevant, or already achieved tasks through the following elements:
        \begin{itemize}
            \item \textbf{High-level task history:} This is a textual summary of the 20 most recent tasks and the 5 most relevant pieces of task knowledge. This historical information is stored in long-term memory by the \textit{Reflector} agent.
            \item \textbf{Total and visited activities:} A list of covered and uncovered activities and the number of visits to each activity, which helps gauge exploration progress and provides details about all activities, including their current state.
            \item \textbf{Initial knowledge:} \textsc{DroidAgent} is initialized with initial knowledge like user virtual profile \textit{Persona}’s information.
        \end{itemize}
        \item \textbf{Acting:} \textsc{DroidAgent} selects a task and provides a prompt containing widget properties (such as Name, text, number of visits, clickable, editable, etc.) to the Actor agent (LLM). The Actor agent is then asked to choose a widget and determine the appropriate action (e.g., click). If the selected action requires one or more parameters, such as text to fill in for an edit action, the Actor LLM is prompted to provide the necessary parameters, using the history of previous prompts and responses related to the current action. In case of an error (e.g., selecting an incompatible option with the widget or providing incorrect parameters like a type mismatch for email or age), a new prompt is sent. This prompt includes the current task's prompt-response history, as well as the error message. After every three iterations, the Actor provides feedback based on the history of the task’s prompts and responses, along with observations. This feedback helps prevent undesirable actions (e.g., exiting the app) or repeating the same action, guiding the Actor to determine the next step for completing the current task successfully.
        \item \textbf{Observing:} After each action is performed, \textsc{DroidAgent} updates its perception of the screen and sends a prompt to the \textsc{Observer} to compare the previous and current GUI states. This information provides feedback to the \textsc{Actor} agent on the effectiveness of the chosen action. Then, \textsc{DroidAgent} proceeds with the next acting step, followed by another observation. This cycle continues until the current task is completed, or the \textsc{Actor} calls the "end task" function.
        \item \textbf{Task reflecting:} After a task execution round concludes—either when the \textsc{Actor} triggers the “end task” function or the maximum action limit is reached, the \textsc{Reflector} is activated. It reviews the task execution history, including self-critique, observations, the current GUI state, and the task’s objective, and generates a concise summary, including a binary success/failure label. The \textit{Reflector}’s purpose is to produce reflections that inform future task planning and improve effectiveness in achieving the overall goal. This reflection process helps preserve useful knowledge from task execution, preventing the system from forgetting important insights and reducing the risk of the other LLM agents straying from their intended purpose or generating irrelevant outputs. Once the reflection is completed, the process restarts at Step 1, continuing until execution times out.
    \end{enumerate}

    The main prompt engineering strategies used are:
    \begin{itemize}
        \item \textbf{Task decomposition:} The \textit{Planner} agent breaks down complex tasks into smaller subtasks, which help create more precise prompts for the LLM.

        \item \textbf{Chain of thought reasoning:} \textsc{DroidAgent} uses step-by-step approach to reason through tasks.

        \item \textbf{Self-reflection:} The \textit{Reflector} agent improves future decisions based on previous experiences.

        \item \textbf{Tool utilisation:} The \textsc{Actor} agent utilises GUI actions and supplies the necessary parameters using this technique.

        \item \textbf{Memory management:} \textsc{DroidAgent} uses both short-term and long-term memory systems during execution, to continue and build upon previous knowledge.

        \item \textbf{Temperature parameter:} It dictates the level of randomness in the model's output. Higher temperatures lead to more varied and diverse responses. Setting the temperature to 0.6 strikes a balance, resulting in outputs that are more factual and reliable \cite{llm_temperature}.
    \end{itemize}

%% file: Sections/3_related_works.tex
\label{sec:RelatedWorks}

Several studies have focused on detecting code smells in mobile apps. Similarly, numerous studies have addressed the dynamic analysis of mobile apps for various purposes. Other research has concentrated on generating execution traces in mobile apps. However, only one study, the \textsc{Dynamics} \cite{Dynamics} tool-based method, has specifically addressed the detection of code smells in mobile apps using dynamic analysis and execution trace generators.

\subsection{Detection of Code Smells in Mobile Apps}
\begin{comment}
parler des méthodes de détection des codes smells statique comme paprika et adoctor et leurs inconvénient.
parler de \textsc{Dynamics} qui est le premier outil de détection qui fait l'analyse dynamique et ses limitations 
Chercher s'il y a d'autre outils de détection des codes smells récents
\end{comment}

The detection of code smells in mobile apps has been widely discussed in the literature. Reimann \textit{et al.} \cite{Reimann2014ATQ} present a catalogue of 30 quality smells for Android, derived from documented practices and developer experiences. These smells affect various aspects like implementation, user interfaces, and database usage, negatively impacting efficiency, user experience, and security. Ghafari \textit{et al.} \cite{Ghafari2017SecuritySI} identified 28 security smells that indicate potential vulnerabilities and developed a static analysis tool for detection. The detection of these smells is based on the presence of certain attributes or method calls, rather than inappropriate code behaviour. The Android community continues to identify new types of smells to improve app quality \cite{SLR_Android_code_smells}, such as the 7 Android-specific behavioural code smells identified by \citet{behavioural_code_smells_impirical_study}. Many approaches capture various features to detect code smells using different analysis techniques in Android apps \cite{SLR_Android_code_smells}. Most studies used for detecting Android-specific behavioural code smells are static-based analysis extensions of \textsc{ADoctor} \cite{ADOCTOR} and \textsc{Paprika} \cite{Paprika}. Both tools use term search and metric computation techniques. Kotlin-specific code smells \cite{Kotlin_bad_smell_detector} have also been explored, but existing tools are limited to detecting only three types of code smells. Machine learning approaches, such as those used to detect wake lock code smells\cite{ML_Wakelocks}, a specific type of behavioural code smell, have shown potential, but require large, balanced datasets. Additionally, Multilayer Perceptron (MLP) \cite{MLP_Wakelocks} models have been applied to detect wake lock smells, though they do not address other resource leaks in mobile apps. 

%Hybrid approaches that integrate static and dynamic analysis are used to detect smells overlooked by other methods, but random-based testing often misses critical execution paths. To improve detection, contextual input generation can be employed to trigger malicious behaviour. Machine learning approaches, such as those used to detect wake lock code smells\cite{ML_Wakelocks}, a specific type of behavioural code smell, have shown potential, but require large, balanced datasets. Additionally, Multilayer Perceptron (MLP) \cite{MLP_Wakelocks} models have been applied to detect wake lock smells, though they do not address other resource leaks in mobile apps. 

 Hybrid approaches \cite{Hybrid1,hybrid2,Jitana} are employed to detect code smells by addressing specific aspects that other methods may overlook. These approaches integrate both static and dynamic analysis to improve the detection process. However, while methods like random-based testing can quickly generate inputs and are efficient, they often fail to explore all possible execution paths, potentially missing malicious code that only executes under certain conditions. To overcome this, Yunmar \textit{et al.} \cite{Hybrid_detection_SLR} suggest using Contextual input generation, which tailors inputs to the application's context, enhancing the chances of triggering and identifying malicious behaviour. This specifically aligns with the intention of our approach.

  %  Concurrently, research utilizing Machine Learning \cite{ML_Wakelocks}, has been conducted to identify wake leak, a specific type of behavioural code smell. However, the effectiveness of this method is contingent upon the acquisition of a substantial volume of balanced data. Additionally, an alternative methodology employing Multilayer Perceptron (MLP) \cite{MLP_Wakelocks}, has been applied to pinpoint eight distinct wake lock code smells. Nevertheless, it is important to note that both of these studies do not address the other resource leak problems that can occur in mobile applications.

\subsection{Dynamic Analysis Used in Mobile Apps}

Dynamic analysis is employed for various purposes in addressing mobile application issues. For instance, \textsc{PredRacer} \cite{PredRacer} leverages dynamic analysis to detect potential data races. This approach helps enhance the search scope and reduce false negatives, while also minimising false positives by incorporating the happen-before relations specific to the Android concurrency model. Dynamic analysis is also widely used in security; for example, JNFuzz-Droid \cite{JNFuzzDroid} applies dynamic analysis for fuzz testing and taint analysis to identify vulnerabilities in Android native code by observing the app's behaviour during execution. Additionally, \textsc{DaPANDA} \cite {DaPanda} introduces a hybrid approach to detect aggressive push notifications in Android apps automatically. Using a guided testing method and instrumenting the Android framework, \textsc{DaPANDA} effectively identifies aggressive notifications%, significantly improving mobile platform user experience
.

\subsection{Execution Trace Generators for Mobile Apps}
    A variety of agents for the generation of execution traces for mobile apps have been developed to enhance the testing process. 
    Random-based agents like \textsc{MonkeyRunner} \cite{monkey_tool} utilise pseudo-random sequences to simulate user interactions, providing a broad, albeit non-targeted, exploration of the AUT.
    On the other hand, model-based agents such as \textsc{Droidbot} \cite{Droidbot} and its advanced iteration, \textsc{DroidBotX} \cite{Droidbotx}, employ a structured approach to testing, using pre-defined models to guide input generation, thereby offering more focused and efficient test coverage, but do not take into account the context of the application, which sometimes causes failure to pass for login activity. \textsc{HumanDroid} \cite{Humanoid} extends this concept by incorporating human-like interactions, aiming to mimic real-world user behaviour more closely by using a deep neural network. However, the training of such models is expensive in time and necessitates a huge amount of data.

    %Moreover, a recently published study \cite{akinotcho2024mobile} indicates that even the most sophisticated GUI exploration techniques achieve only around 30\% coverage of real-world applications. These techniques face several challenges, including difficulties in generating \textit{semantically meaningful inputs} (such as correctly formatting passwords in registration forms), and the order of GUI actions (e.g., the need for sequential actions to bypass the current activity, as demonstrated in our ToyApp).

    The emergence of LLM-based agents marks a significant evolution in this field. Tools like \textsc{GPTDroid} \cite{GPTDroid} and \textsc{Droidbot-GPT} \cite{Droidbot_GPT} leverage the sophisticated capabilities of LLMs to generate \textit{semantically} rich and context-aware interactions with the AUT. \textsc{DroidAgent} \cite{DroidAgent}, in particular, represents a leap forward, autonomously setting and pursuing task goals within the app environment. Its ability to navigate through more activities and perform meaningful tasks specific to the app underlines its potential as a powerful tool in GUI testing. However, the cost-effectiveness of such advanced agents, as exemplified by the operational costs associated with \textsc{DroidAgent}, remains a critical factor for widespread adoption in the industry. In our work, we have developed a version of the \textsc{DroidAgent} tool that uses freely accessible and open-source LLMs.

\subsection{Detection of Code Smells via Dynamic Analysis}
    \textsc{Dynamics} \cite{Dynamics} is the only tool-based method that detects Android-specific behavioural code smells using dynamic analysis. However, this tool suffers from high false negatives due to the low coverage using random-based execution trace generator tools, such as \textsc{MonkeyRunner}. Our tool is also based on the \textsc{Dynamics} method, and to resolve this challenge, we have used an LLM-based trace generator to produce intelligent execution traces.

    There is considerable interest in detecting code smells and conducting dynamic analysis of mobile apps, highlighting the community's focus on these topics. However, there is a notable gap in research concerning the detection of behavioural code smells through dynamic analysis in mobile apps, along with a lack of dedicated tools and methods for this purpose. This gap underscores the relevance of our paper in advancing the state of the art.

\subsection{Code Smells detection using Large Language Models}
    Recent studies have explored the application of LLMs for detecting and correcting code smells. Various LLMs are leveraged for this purpose, including proprietary models such as the GPT-4 series \cite{mesbah2025leveraging} and Gemini \cite{amorimbad}, as well as open-source alternatives like LLaMa \cite{mesbah2025leveraging} and DeepSeek \cite{sadik2025benchmarking}. The \textsc{iSmell} framework \cite{iSmell} utilises a Mixture of Experts approach, integrating multiple code smell detection tools to provide comprehensive analysis and refining LLMs with expert toolset results to facilitate effective refactoring. However, existing methodologies predominantly rely on static analysis, which is insufficient for identifying behavioural code smells and necessitates access to source code. Our work distinguishes itself by employing open-source LLMs within dynamic analysis to generate intelligent execution traces, which addresses these limitations and enhances the detection process without requiring access to source code.
    
    CodaMosa \cite{codamosa} enhances search-based testing by invoking LLM when coverage stagnates, using LLM-generated test seeds as guidance. This helps to achieve higher coverage than LLM-only methods. CodaMosa triggers LLM calls based on coverage plateaus during search, unlike our hybrid method, which uses a runtime trigger based on blocked exploration, switching between fast random actions and semantic LLM-guided execution.

%\subsection{Large Language Models utilisation for Mobile Applications}
%\todo{Large Language Models are used for mobile applications in various tasks, for example, this study \cite{LLMbreakingBarriers} uses LLMs to generate interface code that includes accessibility features, and the results show that LLMs can generate code, but that it sometimes requires adjustments.}

%% file: Sections/4_tool.tex
    \label{sec:SolutionDesign}
    \begin{figure*}[htbp]
    \centering
    \scalebox{0.25}{  
    \includegraphics{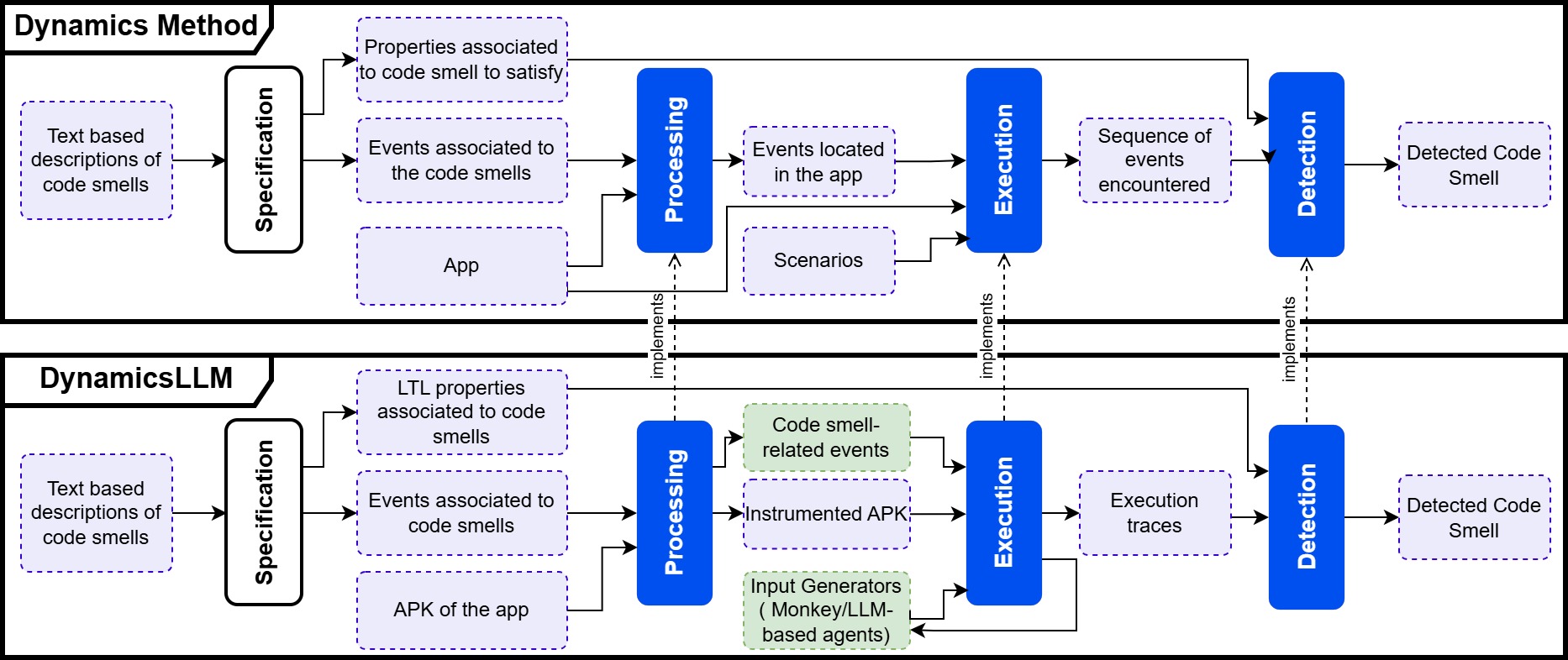}
    }
    \caption{The different steps of the \textsc{Dynamics} Method and the \textsc{DynamicsLLM} Tool. Boxes represent steps, arrows connect the inputs and outputs of each step, described by dotted boxes. Green dotted boxes highlight new or modified elements compared to the \textsc{Dynamics} tool.}
    \Description{DynamicsLLM has four steps: Specification, Processing, Execution, and Detection. The Specification step defines code smells and generates LTL properties along with the events associated with each code smell. The Processing step takes the APK file of the app and the code-smell events as input. It produces two outputs: the code smell-related events and the instrumented APK. The Execution step uses the outputs from Processing along with an input generator to produce execution traces. During execution, DynamicsLLM provides the input generator with information about triggered and newly triggered code smell-related events. The final step, Detection, takes the execution traces as input and verifies the LTL properties to determine whether a code smell has been detected.}
    \label{fig:pipeline}
    \end{figure*}

The \textsc{DynamicsLLM} tool is a concrete implementation of the \textsc{Dynamics} method, which generalises the \textsc{Dynamics} tool by integrating the use of LLMs to intelligently generate execution traces. The different steps of the \textsc{Dynamics} method and the \textsc{DynamicsLLM} tool are illustrated in figure~\ref{fig:pipeline}. \textit{Specification} and \textit{Detection} steps were not modified; only the \textit{Processing} and \textit{Execution} steps are different. We present below each step of the \textsc{DynamicsLLM} tool, explicitly detailing the inputs, outputs, description, processing, and an execution example.

%The \textsc{DynamicsLLM} tool is a concrete implementation of the \textsc{Dynamics} method \cite{Dynamics}, which utilizes LLMs to intelligently generate execution traces. The different steps of the \textsc{Dynamics} method and the \textsc{DynamicsLLM} tool are illustrated in Figure~\ref{fig:pipeline}. In the following, we present each step of the \textsc{DynamicsLLM} tool, explicitly detailing the inputs, the outputs, the description, the implementation, and a running example.

    \subsection{\textbf{Step 1. Specification}}
%    This step follows the same procedure as outlined in the \textsc{Dynamics} tool. In this phase, code smells in Android development are identified by linking specific events (like method calls or instructions) with properties (conditions that indicate the presence of a code smell). By combining these events and properties, behavioural code smells can be represented as conditions that describe inappropriate app behaviours.
    
 %   Each code smell is associated with a property that can be verified, enabling developers to detect the issue as soon as it is confirmed.
         \noindent \textbf{Inputs:} Textual descriptions of mobile code smells found in the literature.
         
         \noindent \textbf{Outputs:} The code smell-related events and the Linear-time Temporal Logic (LTL) property associated with each code smell, when satisfied, indicate the presence of the code smell.
         
        \noindent \textbf{Description:} The code smell-related events are extracted from their descriptions and used to define associated properties. Each event is characterised by specific source code features, such as method calls or code structures, and is linked to values like timestamps or memory metrics.
        
        \noindent \textbf{Implementation:} This step has no implementation. The specification, including the definitions of events and properties, was specified manually by \citet{Dynamics}.
        
        \noindent \textbf{Running Example:}
            \textit{HMU Events} are triggered by calls to the methods of the classes \textit{HashMap}, \textit{SimpleArrayMap} and \textit{ArrayMap} influencing the size of the structures: \textit{new}, \textit{put}, \textit{putAll}, \textit{remove}, \textit{clear}. The values associated with each event are the type of structure, its class, its actual size, and its ID.  %For this example, an HMU is identified when the type is either SimpleArrayMap or ArrayMap with more than 500 elements, or when the type is HashMap with fewer than 500 elements.
    \subsection{\textbf{Step 2. Processing:}}
        \noindent \textbf{Inputs:} The APK file of the app and the events related to the code smells.
        
        \noindent \textbf{Outputs:} The instrumented APK with logging instructions, and a file that contains all the specific events that can cause a code smell.
        
        \noindent \textbf{Description:} In this step, the goal is to instrument the APK of a mobile app by adding logging instructions that differ according to the category the code smell belongs to. Specifically, this involves identifying the relevant events within the APK and inserting instructions that create specific log entries for each event. As a result, when the instrumented app runs, it will generate a trace consisting of a sequence of log entries. In our context, a log entry is represented as a tuple (location, id, event, values), where:
            \begin{itemize}
                \item \textit{location} is the package, class, and method names where the event occurs;
                \item \textit{id} is a sequential identifier that distinguishes several occurrences of the same event in the same method of the same class;
                \item \textit{event} is the keyword associated with a code smell-related event;
                \item \textit{values} contains the values used for the LTL properties of the code smell detection, like the size or the time.
             \end{itemize}
             Unlike \textsc{Dynamics}, \textsc{DynamicsLLM} generates another file that contains information about all the code smell-related events, structured as the tuple (class, method, line number, event).
             
                \begin{figure*}[t]
            \centering
            \begin{lstlisting}[
            emph={new, put, HashMap, release},emphstyle=\textbf, basicstyle=\linespread{1.0}\small, %or \small or \footnotesize etc.
            ]
            HashMap<String, TimePeriod> CACHE = new HashMap<String, TimePeriod>();
            CACHE.put(string, value = new TimePeriod(begin, end));
            \end{lstlisting}
            \caption{Java instruction example.}
            \Description{HashMap<String, TimePeriod> CACHE = new HashMap<String, TimePeriod>();
            CACHE.put(string, value = new TimePeriod(begin, end));}
            \label{fig:examplejava}
        \end{figure*}
        
        \begin{figure*}[t]
          \centering
          
          \begin{lstlisting}[
            emph={hmuadd, HashMap, acquire, release},emphstyle=\textbf, basicstyle=\linespread{1.0}\small, %or \small or \footnotesize etc.
        ]
package.TimePeriodPreference$TimePeriod.java$fromString:0:hmuadd:206399898:1:HashMap\end{lstlisting}
        \caption{Log output associated with the Java instruction.}
        \Description{package.TimePeriodPreference\$TimePeriod.java\$fromString:0:hmuadd:206399898:1:HashMap}
        \label{fig:examplelog}
        \end{figure*}
        \noindent \textbf{Implementation:} This step involves performing static analysis to instrument the code. To achieve this, we added to the \textsc{Dynamics} implementation module a set of code to save the events in a file after the completion of instrumentation.

        \textsc{DynamicsLLM} tool, using \textsc{Soot} \cite{ValleRai2010SootAJ}, processes the representation of each class, method, and instruction within these methods, converting them into Jimple \cite{ValleRai2010SootAJ}, a simplified version of Java source code. Instructions are then analysed to determine if any events are present. If an event is found, instructions to generate log entries are inserted immediately after the event in the APK. Finally, the generated APK must be signed to ensure it can be executed in the next step, which we accomplish using \textsc{Apksigner} \cite{jarsignerurl}.

            \textsc{DynamicsLLM} and \textsc{Dynamics} analyse the bytecode instead of the source code, which has its pros and cons. On the positive side, we do not need access to the source code or any compilation steps. However, translating bytecode into Java or other intermediate languages is challenging due to significant information loss, such as the original line numbers. This is because the type and name of local variables can be difficult to recover, and GOTO statements replace control structures (such as loops and conditionals).

        \noindent \textbf{Running Example:} Figure~\ref{fig:examplejava} shows a call to the \textit{HMU}'s method \textit{put} that add an element to the \textbf{HashMap} structure. This is an \textit{Addition} event for the HMU code smell. In this case, the instrumented app will output the log entry depicted in Figure~\ref{fig:examplelog}. This entry shows the method name \textit{fromString}, the class name \textit{TimePeriodPreference} and the package name \textit{package}
        where this call occurs. It also shows that this is the first event of this type in the method, thanks to the id 0. Finally, it indicates that it is an \textit{Addition} (the \textit{hmuadd} keyword) and that it has operated on the structure of id 206399898, and also the size was 1 and the type of the element is \textbf{HashMap}.

    \subsection{\textbf{Step 3. Execution:}}
    %We run the instrumented app automatically on an Android emulator using our trace generator to monitor its actions. As actions are executed, multiple log entries are generated, collectively forming an execution trace. Unlike the \textsc{Dynamics} tool, \textsc{DynamicsLLM} does not rely solely on traditional trace generators; instead, it can leverage LLM-based generators.

    \noindent \textbf{Inputs:} The instrumented APK, the code smell-related events file, and the LLM-based input generator.
    
    \noindent \textbf{Outputs:}
    The execution traces are obtained after running the instrumented app with an LLM-based trace generator.
    
    \noindent \textbf{Description:} This step involves executing the instrumented app on a real or virtual device, using an LLM-based agent with different configurations (such as APK file, event files, allowed execution time, device serial, output directory) to produce execution traces. 
    This agent is a program that simulates user interactions by generating inputs for the app, such as clicking a specific button, entering text in a field, or navigating backwards. 
    This step is crucial for the \textsc{DynamicsLLM} tool, as the generated traces will facilitate the detection of code smells. Each time execution reaches a logging instruction, a corresponding log entry is produced. Consequently, the execution trace comprises all these log entries, reflecting the events associated with the detected code smells. The key challenge is to achieve comprehensive coverage of the instrumented code during runtime, capturing as many events as possible to maximise the potential for detecting code smells.
    
    \noindent \textbf{Implementation:}
            We ultimately chose \textsc{DroidAgent} due to its advanced LLM-based capabilities. The LLM-based tools offer enhanced intelligence, require minimal training, and are designed to understand application contexts, allowing them to plan action sequences to achieve their objectives dynamically. We decided against \textsc{GPTDroid} because its code is not publicly available \cite{DroidAgent} and it functions solely as an actor agent. In contrast, \textsc{DroidAgent} operates as a task-based agent, planning steps during execution based on assigned tasks.
            
            The main modifications applied to \textsc{DroidAgent} are outlined below, with examples of the updated prompts included in the replication package \cite{replicationPackage}.

            \begin{itemize}
                \item \textbf{Using Open-Source LLMs:} All our LLM actors are powered by freely accessible and open-source language models. We updated the model module in \textsc{DroidAgent} to send prompts to LLMs running either locally or on a server via Ollama \cite{ollama2024}, thus removing the limitation to the GPT model found in the original version.
                \item \textbf{Redefining the ultimate objective:} 
                For the Planner agent, the new objective was modified from the default "[PERSONA]’s ultimate goal is to visit as many pages as possible and try their core functionalities" to "[PERSONA]’s ultimate goal is to trigger code smell-related events as much as possible".

                \item \textbf{Extend the initial knowledge:} We have extended the initial knowledge for the \textit{Planner} agent by including a list of potential code smells, obtained from the instrumentation, formatted as follows: "(Type of code smell, class, method): [... ]". This list is also used in the observing step to know the triggered events.

                \item \textbf{Observe prompt:} At each observation step, the prompt sent to the \textit{Observer} includes both the visible activity changes and detailed information about triggered code smell events. Specifically, it reports: (i) the number of code smell-related events since the last observation, (ii) the cumulative number of triggered events, and (iii) the total number of code smell-related events present in the app.

            \end{itemize}

            Our solution is a fully automated, we use \textsc{DynamicsLLM} tool on \textsc{Google Android emulator} \cite{androidstudio2024} to generate the traces that will be used for detection. We used the \textsc{ADB} \cite{android_adb} command-line tool to install the APK, \textsc{UIAutomator2} \cite{uiautomator2} for extracting the view hierarchy, and \textsc{logcat} \cite{android_sdk_tools} command-line tool to retrieve the log of the device, which contains the generated execution traces, during execution. We filter those logs to select only the logs that match our log format. So, there is no filtering on the sequences generated.
            
        \noindent \textbf{Running Example:} Figure~\ref{fig:lambda} presents an excerpt from a \textbf{trace generated} during the execution aimed at detecting the \textbf{HMU} code smell. This excerpt is derived from the running example provided in Step 2. The trace shows one \textbf{implementation} and four instances of adding the identity $206399898$ to the \textbf{HashMap}. It indicates that the maximum size attempted is 4, and throughout the execution, the \textbf{HashMap} was used for a small set, which constitutes an HMU code smell.

    \subsection{\textbf{Step 4. Detection:}}

        \begin{figure*}[tb]
              \centering
              \begin{lstlisting}[
                emph={g, h, acquire, release},emphstyle=\textbf, basicstyle=\linespread{1.0}\small, %or \small or \footnotesize etc.
            ]
            package.TimePeriodPreference.java$<clinit>:0:hmuimpl:206399898:0:HashMap
            package.TimePeriodPreference.java$fromString:0:hmuadd:206399898:1:HashMap
            package.TimePeriodPreference.java$fromString:0:hmuadd:206399898:2:HashMap
            package.TimePeriodPreference.java$fromString:0:hmuadd:206399898:3:HashMap
            package.TimePeriodPreference.java$fromString:0:hmuadd:206399898:4:HashMap
            \end{lstlisting}
            \caption{Snippet of an execution trace.}
            \label{fig:lambda}
            \Description{The listing presents an excerpt of an execution trace generated during the dynamic analysis of the application. Each entry corresponds to a runtime event associated with a specific method execution. The trace records the class and method name, the execution context, a unique object identifier, the invocation order, and the object type involved in the operation. In this example, the static initializer (\$<clinit>) of the TimePeriodPreference class creates a HashMap instance, followed by multiple invocations of the fromString method that progressively add elements to the same HashMap object. The repeated identifier indicates that all operations are performed on the same runtime instance, enabling the reconstruction of the object’s behavioral evolution throughout execution.}
            \end{figure*}
            
        \begin{figure*}[tb]
              \centering
              \begin{lstlisting}[
                emph={HashMap, hmuimpl,\,},emphstyle=\textbf, basicstyle=\linespread{1.0}\small, %or \small or \footnotesize etc.
            ]
            package.apk,package,package.TimePeriodPreference$TimePeriod.java,<clinit>,HashMap,4
            07:52:02.035,package.TimePeriodPreference.java$<clinit>,0,hmuimpl
            \end{lstlisting}
            \caption{Code smell detection format.}
            \Description{package.apk,package,package.TimePeriodPreference\$TimePeriod.java,<clinit>,HashMap,4
            07:52:02.035,package.TimePeriodPreference.java\$<clinit>,0,hmuimpl}
            \label{fig:event_log}
            \end{figure*} 
            
    As with the \textsc{Dynamics} tool, \textsc{DynamicsLLM} will detect code smells from the app's complete execution trace.
    
        \noindent \textbf{Inputs:} A set of execution traces and the LTL properties associated with the code smells.
        
        \noindent \textbf{Outputs:} The code smells detected during the execution are grouped in files by type.
        
        \noindent \textbf{Description:} This step involves analysing execution traces using runtime monitoring to identify code smells by examining event sequences and verifying code-smell properties. Only events associated with this specific instance or method call should be used for the verification, ensuring that properties are not checked using events from different objects or calls.
        
        \noindent \textbf{Implementation:}
            For this step, we utilised a Java module that performs runtime monitoring in our tool using the \textsc{SOOT} framework. Alternatively, we could also employ \textsc{BeepBeep 3}, which is integrated into the \textsc{Dynamics} tool \cite{Hall2017EventSP}. In both cases, we specify a branch for each code smell to verify the corresponding property. Once the entire trace is processed, we identify the detected code smells. Each code smell will be added to the file of the same type of events as a tuple, which contains all the necessary information; the structure differs from one type of code smell to another.

        \noindent \textbf{Running Example:} For the code smell detected in Figure \ref{fig:lambda}, a new line is added to the HMU events file as shown in Figure \ref{fig:event_log}. The entries include the APK name, package, file containing the class, and method (in this case, it is \textbf{clinit}, the constructor).
        The details included depend on the code smell type. In the case of HashMap, for example, there are two other informations included: the \textit{structure Type} (HashMap, ArrayMap, or SimpleArrayMap), and also the \textit{maximum size} attempted during execution, which is 4.
            % For validation purposes, we also save the events found during the execution trace in a file, structured as tuples containing the timestamp of the triggered event, the method where the event occurred, the specific line of code to differentiate between various events within the same method, and the event type. 
    \subsection{The hybrid approach}
    This work presents an innovative method for the \textit{Execution} step that merges the LLM approach with a random approach. Instead of relying on traditional time-based methods, we utilise \textsc{MonkeyRunner} to explore mobile applications. When \textsc{DynamicsLLM} detects that \textsc{MonkeyRunner} is 'blocked' (i.e., no longer effective in identifying new code smell-related events in the last 5 minutes), we switch to \textsc{DroidAgent} for 5 steps to 'unblock' \textsc{MonkeyRunner}. We then reuse \textsc{MonkeyRunner} until it is 'blocked' again. The LLM agent (\textsc{DroidAgent}) produces high-quality actions but operates slowly, while the random agent (\textsc{MonkeyRunner}) is very fast but lacks intelligence. This method leverages the strengths of both techniques to maximise coverage.

%% file: Sections/5_evaluation.tex
\label{sec:Evaluation}
    
In this section, we will conduct a comprehensive analysis of \textsc{DynamicsLLM} that leverages LLMs against the latest state-of-the-art \textsc{Dynamics} tool. %\FA{Dire que \textsc{Dynamics} est le state of the art}.
%This comparative study aims to highlight the capabilities and performance metrics of both tools such as precision, recall, number of code smells detected and the coverage of the code smell-related event. The results will provide valuable insights into the effectiveness of LLMs in generating an intelligent execution trace compared to traditional methods and will serve as a guide for future improvements and developments.

%\FA{Ajouter une petite intro qui dit qu'on va évaluer notre outil uniquement en se comparant a Dynamics (qui utilise monkey pour la génération des traces)}

\subsection{Research Questions}
    We address the following three research questions:
    \begin{itemize}
        \item \textbf{$\boldsymbol{RQ_1}$: Does the intelligent traces generated by \textsc{DynamicsLLM} allows the detection of behavioural code smells?} This question investigates the effectiveness of \textsc{DynamicsLLM} for the detection of behavioural code smells in terms of precision and recall.
        % répondre par le tableau de precision et de rappel des experimentations: hybrid 1 heure et mistral 100 action timeout 3 heures
        
        \item \textbf{$\boldsymbol{RQ_2}$: Does \textsc{DynamicsLLM} cover more code smell-related events compared to the state-of-the-art tool?} This question investigates the coverage of \textsc{DynamicsLLM} on triggering the code smell-related events.
        % Je réponds à cette question par les graphes de nombre d'évènements en function du temps et d'actions

        \item \textbf{$\boldsymbol{RQ_3}$:What is the contribution of LLMs to overall effectiveness?} This research question aims to respond whether it is still worthwhile to use \textsc{Dynamics} and under which conditions or scenarios \textsc{DynamicsLLM} offer significant advantages.
        % Utiliser le graphe de trade-off et le tableau 3 et la figure 8, en expliquant quels sont les patterns où les LLMs sont mieux et ou c'est MonkeyRunner est mieux, et expliquer les patterns qui sont bloquants.
        
    \end{itemize}

\subsection{Experimental Setup}
    %To address these research questions, we executed both \textsc{Dynamics} executed using \textsc{MonkeyRunner} in 1 hour, and \textsc{DynamicsLLM}, using two configurations: \textsc{100\% LLM} executing at most 100 actions with a timeout of 3 hours, and \textsc{Hybrid} executed for 1 hour.
    To address these research questions, we executed these configurations:
    \begin{itemize}
        \item \textsc{Dynamics} executed using \textsc{MonkeyRunner} in 1 hour.
        \item \textsc{DynamicsLLM} using \textsc{100\% LLM} executing at most 100 actions with a timeout of 3 hours.
        \item \textsc{DynamicsLLM} using \textsc{Hybrid} executed for 1 hour.
    \end{itemize}
    We will denote the first configuration as \textsc{MonkeyRunner}, the second as \textsc{100\% LLM}, and the third as \textsc{Hybrid}.
    
    The dataset is a subset of the \textbf{real-world} open-source F-Droid dataset \cite{FDroid}. Applications that could not be built or executed using the selected Android SDK version and \textsc{MonkeyRunner} were excluded, resulting in a final set of 333{} applications. Among them, 136{} applications were manually annotated by \citet{Dynamics} for validation, representing a stratified sample that is statistically significant at the 95\% confidence level with a 10\% margin of error\cite{behavioural_code_smells_impirical_study}. The collected metrics are:
    \begin{itemize}
        \item \textbf{Precision:} The proportion of true positives(TP) among all detected positives(D) (i.e., detected code smells), calculated as $\frac{\lvert TP \rvert}{\lvert D \rvert}$.
        
        \item \textbf{Recall}: The proportion of true positives(TP) among the sum of true positives and false negatives(FN), calculated as $\frac{\lvert TP \rvert}{\lvert TP \cup FN \rvert}$.
        
        \item \textbf{Number of code smell-related events covered:} The number of instrumented lines in the code that are reached during trace generation, counted only once per line.
    \end{itemize}

    %\FA{Notons que les True Positive proviennent de ... (des anotations a la main + ce que les outils on détecté ?)}
    % The validation dataset was constructed from the manual annotations of Dimitri \textit{et al.} \cite{behavioural_code_smells_impirical_study}.

\subsection{Experiment Environment}
\begin{figure}[htbp]
        \centering
        \scalebox{0.6}{
            \includegraphics{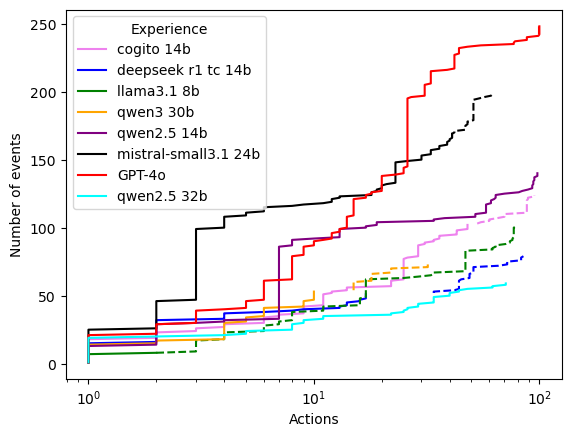}
        }
        \caption{Number of code smell-related events covered by each model. The dotted line indicates that the timeout was reached for certain executions.}
        \Description{Comparison of large language models showing that GPT-4o triggers the highest number of code smell-related events (250 events), followed by Mistral-Small-3.1-24B with approximately 200 events. Other models trigger fewer events.}

        \label{fig:LLMs_results}
    \end{figure}

    All experiments were conducted on a 64-bit Ubuntu 22.04 machine using the Google Android Emulator \cite{androidstudio2024} (API 35), along with Android SDK Command Line Tools 16.0 \cite{android_sdk_tools}. We chose this version because we observed that \textsc{DroidAgent} could not fill in text fields in the previous versions, due to an incompatibility with the \textsc{Droidbot} package version. 
    \noindent The setup is connected to the Ollama RESTful API server \cite{ollama2024}, which manages models executed on a GPU server to serve LLMs. We selected \textbf{mistral-small3.1 24b} based on a preliminary analysis of 15 representative Android apps. The results, illustrated in Figure~\ref{fig:LLMs_results}, show the number of code smell-related events triggered by our enhanced \textsc{DroidAgent} as a function of the number of actions performed. Notably, GPT-4o achieved the highest event count with fewer actions, indicating greater coverage efficiency. \textbf{Mistral-small3.1 24b} also demonstrated competitive performance, highlighting its capacity to take impactful steps early in the execution. All of these models were able to pass our LoginApp, except LLaMa3.1 8b, likely because it is not a reasoning model. We chose the Mistral-small model because our goal was to use an open-source LLM. Therefore, GPT-4o is included solely as a baseline for comparison.
    
\subsection{Results}
\par In this section, each research question is answered separately.
%we will study and answer the research questions case by case.

\subsubsection{RQ1} \textit{Does the intelligent traces generated by \textsc{DynamicsLLM} allow the detection of behavioural code smells?}

    To answer this question, we evaluated the efficiency of all tools using precision and recall as metrics.
 
    \textsc{DynamicsLLM} tool using \textsc{100\% LLM} and \textsc{Hybrid} configurations demonstrated 100\% precision for all types of code smells, meaning that all the detected anomalies were \textbf{accurate}.

    Tables~\ref{tab:recall} present recall metrics for both \textsc{DynamicsLLM} across the tested applications, differentiated by various code smell types. 
    Detection of DW and IOD was more modest across all tools. For DW, this is due to low coverage of such instructions, but for the IOD, it is attributed to the fact that \textsc{DynamicsLLM} does not saturate the activity through sustained high-speed interactions over time, exhibiting usage patterns that differ significantly from those of human users.
    However, for HMU and NLMR smells, all configurations achieved high recall, with 73\% for HMU and 98\% for NLMR.
    \begin{table}
\renewcommand{\arraystretch}{1}
\caption{Recall for code smell detection}
  \label{tab:recall}
  \centering
\begin{tabular}{lll}
\toprule
Code Smell & \textsc{Hybrid} & \textsc{100\% LLM} \\
\midrule
DW & 1 / 8 (12\%) & 2 / 8 (25\%) \\
HP & 8 / 13 (61\%) & 8 / 13 (61\%) \\
IOD & 2 / 7 (28\%) & 3 / 7 (42\%) \\
NLMR & 1033 / 1057 (97\%) & 1033 / 1057 (97\%) \\
HMU & 154 / 210 (73\%) & 155 / 210 (73\%) \\
\bottomrule
\end{tabular}
\end{table}

\begin{quote}
\noindent\textbf{$\boldsymbol{RQ_1}$}: 
\textsc{DynamicsLLM}, using both the 
\textsc{100\% LLM} and \textsc{Hybrid} configurations, is capable of detecting 
Android-specific behavioural code smells, demonstrated by its precision (100\%) 
across all code smell types and competitive recall scores.
\end{quote}

\subsubsection{RQ2} \textit{Does \textsc{DynamicsLLM} cover more code smell-related events compared to the state-of-the-art tool?}
    
    To assess the potential of LLM-based approaches, we compared the code smell-related events covered by each tool. The significance of the results was evaluated using the McNemar test, and the detailed outcomes are provided in the replication package\cite{replicationPackage}.

\begin{figure*}[htbp]
    \centering
    \scalebox{0.72}{
        \includegraphics{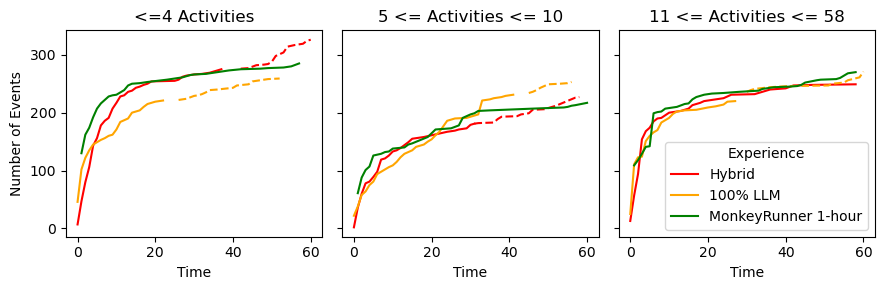}
    }
    
    \caption{Code smell-related events covered by each tool. The dotted line indicates that certain executions ended before.}
    \Description{Coverage results grouped into three categories based on the number of activities per app: apps with at most four activities, apps with five to ten activities, and apps with eleven or more activities. For the first category, Hybrid triggers the highest number of events (over 326), followed by MonkeyRunner (around 285) and 100\% LLM (259). In the second category, 100\% LLM triggers the most events (253), followed by Hybrid (227) and MonkeyRunner (217). In the third category, the three configurations trigger a similar number of events, with MonkeyRunner (270), 100\% LLM (271), and Hybrid (249).}

    \label{fig:uniques_by_time}
\end{figure*}
    Figure~\ref{fig:uniques_by_time} shows the number of code smell-related events triggered over time by different approaches across three app categories: 211 apps with four or fewer activities, 78 apps with between five and ten activities, and 44 apps with more than ten activities.

Within a 1-hour timeout, the \textsc{Hybrid} configuration outperforms the others for apps with fewer activities, demonstrating rapid coverage and achieving improvements of 14.4\% and 25.9\% over \textsc{MonkeyRunner} and \textsc{100\% LLM}, respectively. This category represents 63\% of the dataset. The McNemar test results confirm that the differences between \textsc{Hybrid} and the other configurations are statistically significant. These results suggest that combining strategies is highly effective.

For apps containing between five and ten activities, results show that \textsc{100\% LLM} achieves 11.4\% and 16.5\% higher coverage than \textsc{Hybrid} and \textsc{MonkeyRunner}, respectively. This demonstrates the strength of LLM-based models for this category; the reasons for this will be further discussed in the answer to \textit{RQ3}.

The McNemar test results also indicate significant differences between the coverage achieved by \textsc{100\% LLM} and the other approaches for apps with between 5 and 10 activities, as well as for those with more than 10 activities, where the final performances are competitive across the three configurations.

Across all three app categories, we observe that \textsc{100\% LLM} initially lags behind \textsc{MonkeyRunner} during the first few minutes, but it gradually narrows the gap over time. This slower initial performance occurs because each proposed action requires 4 steps, each involving one or more interactions with the LLM, which is time-consuming. %Additionally, the LLM occasionally produces responses in the wrong format—likely due to hallucination—necessitating re-prompting until the correct format is obtained.
    %
    % indicate that combining both exploration techniques leads to significantly higher coverage in a shorter timeframe.

\begin{figure}[b]
    \centering
    \scalebox{0.45}{
        \includegraphics{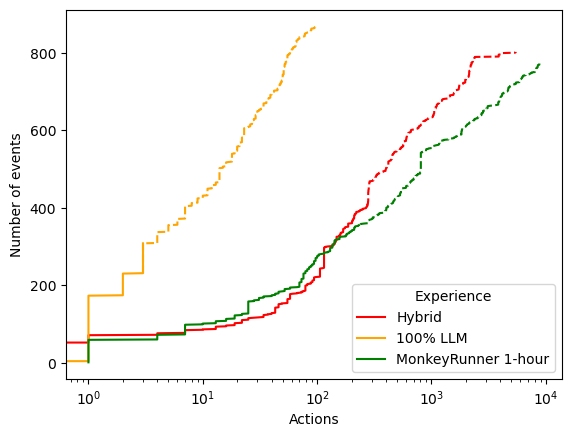}
    }
    \caption{Number of code smell-related events covered by each tool by the number of actions. The dotted line indicates that the timeout was reached for certain executions.}
    \Description{Total number of triggered events for each configuration: 100\% LLM with 868 events, Hybrid with 802 events, and MonkeyRunner with 772 events.}

    \label{fig:uniques_by_actions}
\end{figure}

    Figure~\ref{fig:uniques_by_actions} presents the number of code smell-related events triggered by each tool, plotted against the number of actions. This metric is particularly relevant given the rapid advancements in LLMs. Furthermore, some apps may be more demanding in terms of the number of actions required to complete a test rather than the time taken. As the number of test actions increases, the complexity of debugging also grows, making failures more difficult to trace, reproduce, and resolve. 

    The \textsc{100\% LLM} configuration demonstrates excellent efficiency, achieving around 868 events with only 100 actions (3 times compared to \textsc{Hybrid} and \textsc{MonkeyRunner}). This further confirms the effectiveness of LLMs. %, even without hybridisation.
    The \textsc{Hybrid} configuration also stands out, discovering the highest number of events while requiring significantly fewer actions than \textsc{MonkeyRunner}. The similarity in early-stage results between \textsc{Hybrid} and \textsc{MonkeyRunner} is due to \textsc{Hybrid} initially relying on \textsc{MonkeyRunner} during the first 5 minutes. These findings demonstrate that \textsc{DynamicsLLM} not only discovers more events but does so with far fewer actions. This is because LLM-based exploration can reason about the interface and select more intelligent actions, thereby reducing redundancy.

\begin{quote}
\noindent \textbf{$\boldsymbol{RQ_2}$}: \textsc{DynamicsLLM} using \textsc{100\% LLM} covers more code smell-related events within a fixed number of actions, while the \textsc{Hybrid} performs better in covering events within a fixed time limit for mobile apps that have few activities.
\end{quote}

\subsubsection{RQ3} \textit{What is the contribution of LLMs to overall effectiveness?}

    This question, subdivided into specific research sub-questions, was examined using both quantitative and qualitative analyses.

    First, \textbf{are the code smell-related events covered by \textsc{Dynamics} simply a subset of those captured by \textsc{DynamicsLLM}?}
    To answer this question, we compared the events detected by each tool, identifying which were uniquely captured by one tool or the other (see Table~\ref{tab:table_code_smells_llm}). Our analysis reveals that \textsc{DynamicsLLM} detected 309 and 224 code smell-related events-using \textsc{100\% LLM} and \textsc{Hybrid}, respectively—that \textsc{Dynamics} missed. Conversely, \textsc{Dynamics} identified 213 and 194 events, respectively, that were not captured by \textsc{DynamicsLLM} under the same configurations.
    Additionally, we performed \textbf{McNemar} tests, which showed a statistically significant difference in coverage between \textsc{DynamicsLLM} (using \textsc{100\% LLM}) and \textsc{Dynamics} (p-value = 0.000026), indicating that the observed differences are unlikely to be due to randomness.
       \begin{table}[]
    \renewcommand{\arraystretch}{1}
    \caption{Code Smell-related Events contingency table between \textsc{DynamicsLLM} configurations and \textsc{MonkeyRunner}}
    \label{tab:table_code_smells_llm}
    \centering
\renewcommand{\arraystretch}{0.95}
    \begin{tabular}{lll}
\toprule
 & In \textsc{MonkeyRunner} & Not in \textsc{MonkeyRunner} \\
\midrule
In \textsc{100\% LLM} & 559 & 309 \\
Not in \textsc{100\% LLM} & 213 & 3356 \\
\midrule
In \textsc{Hybrid} & 578 & 224 \\
Not in \textsc{Hybrid} & 194 & 3441 \\
\bottomrule
\end{tabular}
    \end{table}
    
    \textbf{Why does \textsc{Dynamics} fail to trigger certain code smell-related events, even after executing up to 100× and 4× more actions than \textsc{DynamicsLLM} using \textsc{100\% LLM} and \textsc{Hybrid}, respectively?} This is due to the core random exploration strategy, which struggles to reach specific activities and behaviours. In some cases, the order of GUI actions is crucial. For example, we created a toy app with two activities: the first contains a "Next" button that must be clicked 20 times in succession to reach the second activity, where the code smell resides. Even after one hour of execution, \textsc{Dynamics} failed to access this second activity.
    
    Moreover, semantically meaningful input is essential; it is particularly challenging—if not impossible—for random approaches to navigate activities such as login and registration, which require precise form completion. To further illustrate this limitation, we developed an additional mobile application containing Register, Login, and Home activities. Consistent with previous results, \textsc{Dynamics} was unable to reach the Home activity, which includes a behavioural code smell. In contrast, in both cases, the \textsc{DynamicsLLM} tool configured with \textsc{100\% LLM} successfully triggered them.
    
    Finally, \textbf{what are the limitations of using LLMs in dynamic analysis?} Based on our observations, using \textsc{100\% LLM} occasionally has difficulty achieving effective test coverage. For example, in a piano application with numerous colored buttons, none of the UI elements contained textual descriptions. In the absence of such a textual context, it became challenging for the model to select appropriate actions, as its decision-making relies heavily on semantic cues; only 10 actions were taken in 3 hours of testing. 
    
    In another case, the UI of some applications included a long vertical list of widgets, resulting in an overly lengthy description that exceeded the model's input context limit. Consequently, the truncated context omitted widgets located further down the list, preventing the model from reasoning about those components.

    Additionally, interacting with LLMs is time-consuming. \textsc{DynamicsLLM} fails in some complex tasks. Certain games require identifying an object in the UI, and it is limited to only 3 attempts within a short time frame.
    Using LLMs is also resource-consuming. This is not a major issue, as ongoing scientific and industrial efforts are producing more efficient models with lower hardware requirements.

\begin{quote}
\noindent\textbf{$\boldsymbol{RQ_3}$}: The LLM-based approach overcomes traditional limitations of execution trace generators by producing semantically meaningful and logically ordered actions. However, its effectiveness is constrained by the quality and completeness of the contextual UI description.
\end{quote}

\subsection{Threats to Validity}
\label{sec:ThreatsToValidity}
While this study provides valuable insights into the effectiveness of \textsc{DynamicsLLM} for detecting behavioural code smells, several threats to validity must be considered:

\noindent \textbf{Internal Validity.} 
The inherent randomness of LLMs and \textsc{MonkeyRunner} can influence the results. Since multiple runs with different trace-generation strategies can yield varied outcomes, the lack of repeated experiments due to time constraints may introduce biases in our findings.

\noindent \textbf{External Validity.} 
The generalisability of our results to other datasets, app types, and platforms is a key concern. Although we evaluated \textsc{DynamicsLLM} on 333{} apps from the F-DROID dataset, which provides a broad range of application types, it may not fully reflect the diversity of apps on platforms like the Google Play Store. 
%Additionally, F-DROID apps vary in maturity and code quality, with more mature apps potentially exhibiting fewer detectable code smells compared to those in early development, limiting the broader applicability of our conclusions.

%\par \textbf{Construct Validity.}  
%Our study employed common evaluation metrics—precision, recall, and code smell-related events covered, which are effective but may not capture all aspects of behavioural code smells. Certain code smells, especially subtle or intermittent ones, may not be adequately represented by a simple event count. Context-dependent code smells may require a more nuanced approach to assessment that goes beyond traditional metrics.

\noindent \textbf{Repeatability Validity.} 
A strength of this study is the repeatability of our results, as all tools and datasets used are open-source and publicly available in \cite{replicationPackage}. \textsc{DynamicsLLM} and \textsc{Dynamics}, along with the execution traces and results, are reproducible, allowing future researchers and practitioners to validate and extend our findings.

%% file: Sections/6_conclusion.tex
\label{sec:Conclusion}
In this paper, we present \textsc{DynamicsLLM}, a novel tool that uses LLMs to intelligently generate execution traces for detecting behavioural code smells in Android mobile applications. Our tool enhances the effectiveness of dynamic analysis-based methods by addressing issues with traditional trace generators, which typically lead to low coverage and a high false negative rate.

We evaluate \textsc{DynamicsLLM} using a set of 333{} real-world mobile applications to compare its performance with the state-of-the-art \textsc{Dynamics} tool. Our results show that, when restricted to a fixed number of actions, \textsc{DynamicsLLM} using the \textsc{100\% LLM} approach significantly outperforms \textsc{Dynamics} in covering code smell–related events. However, when the time limit is set to one hour, our proposed \textsc{Hybrid} method demonstrates faster and superior coverage, particularly for mobile applications containing few activities, while achieving competitive coverage for other categories.

Additionally, our results highlighted the potential of LLM-based approaches, as \textsc{DynamicsLLM} was able to trigger 12\% more code smell-related events than \textsc{Dynamics} tool, underscoring the advantages of intelligent execution trace generation. Integrating LLMs can be beneficial for other research challenges involving dynamic analysis. This includes mobile application issues such as data race detection, fuzz testing, and taint analysis in mobile security and extends to other platforms. Although LLMs remain relatively slow, the field is advancing rapidly, and faster models requiring fewer computational resources are likely to emerge soon.

In conclusion, this work presents a promising avenue for future research. Subsequent efforts could focus on developing faster LLM-based methods by integrating them with complementary approaches. In particular, Vision-Language Models offer valuable potential for analyzing non-textual UI elements and improving the interpretation of visual feedback.